\documentclass[aip,subeqns.floatfix,twocolumn,amsmath,showpacs]{revtex4}
\usepackage{graphicx,epsfig}
\usepackage{subfigure}
\begin{document}
\title{Transition from amplitude to oscillation death in a network of oscillators}
\author{ Mauparna Nandan$^{1,2}$, C.R.Hens$^3$, Pinaki Pal$^2$, Syamal K.Dana$^3$ }
\affiliation{$^1$Dr. B. C. Roy Engineering College, Durgapur 713206, India}
\affiliation{$^2$Department of Mathematics, National Institute of Technology, Durgapur 713209, India}
\affiliation{$^3$CSIR-Indian Institute of Chemical Biology, Kolkata 700032, India}
 %\author{AIP Journal Program}%
%\email{tex@aip.org}
%\affiliation{American Institute of Physics\\Suite 1NO1, 2 Huntington Quadrangle\\Melville, New York 11747-4502, USA}%
%\date{March 2010}%
%\revised{August 2010}%
\begin{abstract}
We report a transition from a homogeneous steady state (HSS) to inhomogeneous steady states (IHSSs) in a network of globally coupled identical oscillators.  We perturb a synchronized population of oscillators in the network with a few local negative  or repulsive mean field links.  The whole population splits into two clusters for a certain number of  repulsive mean field links and a range of coupling strength. For further increase of the strength of interaction these clusters collapse  into a HSS followed by a transition to IHSSs where all the oscillators populate either of the two stable steady states.  We analytically determine the origin of HSS and its transition to IHSS in relation to the number of  repulsive mean-field links and the strength of interaction using a reductionism approach to the model network. We verify the results  with numerical examples of the paradigmatic Landau-Stuart limit cycle system and the chaotic Rössler oscillator as dynamical nodes. During the transition from HSS to IHSSs, the network follows the Turing type symmetry breaking pitchfork or transcritical bifurcation depending upon the system dynamics.
 \end{abstract}
%\pacs{ 05.45.Xt, 05.45.Gg}
\maketitle
\begin{quotation}
{\bf Quenching of oscillation by coupling of oscillators is a well known fact; it has two different manifestations, HSS or amplitude death (AD) and IHSSs or oscillation death (OD). It is recently reported that a transition from a HSS  to IHSSs  in two diffusively oscillatory units occurs via the Turing type pitchfork bifurcation, in a similar fashion, as noted in a spatio-temporal medium. It attracted attention of many researchers in recent time. However, most of the previous works deals with Landau-Stuart limit cycle systems. We extend the premises of the phenomenon of transition from HSS to IHSS to a network of oscillators, and  furthermore use both limit cycle and chaotic systems. In a synchronized network of globally coupled oscillators, the HSS emerges due to additional repulsive interaction or link as it may arise with time due to an evolving fault or defect and incurs a transition to IHSSs  due to the increasing number of repulsive links as a process of spreading of the fault in the network. We explore analytically and numerically as well the regions of HSS and IHSS and demarcate their boundaries in a parameter space of coupling strength and number of repulsive links and, the bifurcation processes during these transitions.}
\end {quotation}
\section{Introduction}
An emergent and intriguing phenomenon in coupled nonlinear dynamical systems is the oscillation quenching \cite{G Saxena:Phys Rep_2012,A Koseska:Phys Rep_2013} due to coupling interaction.   
This suppression of  oscillation was first observed as an effect of a large parameter mismatch in diffusively coupled oscillators ~\cite{Kopell:Physd_1990,PC Matthews:PRL_1990} whereas Reddy {\it et al} ~\cite{DV Ramana:PRL_1998} later observed this in identical oscillators with time-delayed interactions. Such a cessation of oscillation in dynamical systems was also found to appear for different other forms of coupling, namely, dynamic coupling~\cite{K Konishi:PRE_2003}, conjugate coupling~\cite{R Karnatak:PRE_2007}, and for introducing a damping effect by the environment~\cite{Resmi:Pre_2011} or a repulsive mean field interaction~\cite{CR Hens:PRE_2013, CR Hens:PRE_2014}. 
\par There are two distinct manifestations of the quenching effect, namely, amplitude death (AD) and oscillation death (OD) which are distinguished ~\cite{A Koseska:Phys Rep_2013} by their dynamical properties.
Amplitude death (AD) appears in a coupled system  via the reverse Hopf bifurcation (HB) \cite{N Kopell:PhyscaD_1990} and induces a homogeneous behavior in a coupled system when all the oscillators populate a stable homogeneous steady state (HSS)~\cite{A Koseska:Phys Rep_2013,A Koseska:PRL_2013}. 
In the OD state, the  coupled systems form at least two groups when each group populates different stable steady states known as stable inhomogeneous steady states (IHSS) \cite{K Bar-eli:PhysicaD_1984, Koseska:PRE_2007,W Liu:Chaos_2012}. 
More recently, the borderline between the AD and the OD has been exemplified \cite{A Koseska:PRL_2013,CR Hens:PRE_2013,W Zou:PRE_2013, Banerjee} by showing a  transition from a HSS (AD) to the IHSSs (OD) by a symmetry-breaking instability in two coupled oscillators. 
Such a transition was originally introduced by Turing in a seminal paper~\cite{A Turing:PTRSL_1952} where  he explained the origin of stable patterns in a homogeneous medium due to a growing instability created by a symmetry-breaking diffusion process intrinsic to the system.  
Koseska {\it et al} \cite{A Koseska:PRL_2013} exte
nded this fundamental process of transition from HSS (AD) to IHSS (OD)  via the Turing type symmetry breaking pitchfork (PF) bifurcation to a simple model of two diffusively coupled limit cycle  Landau-Stuart (LS) system~\cite{A Koseska:PRL_2013} with a parameter mismatch. Subsequently, this transition was reported \cite{W Zou:PRE_2013} in two LS systems for different other types of diffusive coupling. In another context, Hens {\it et al} \cite{CR Hens:PRE_2013} reported this transition in two diffusively coupled synchronized oscillators, both limit cycle and chaotic, for additional  repulsive mean-field interaction. 
\par We attempt here to extend the previous results \cite{CR Hens:PRE_2013, CR Hens:PRE_2014} to a network of oscillators and to explore the origin of AD and its transition to OD in a network of oscillators. 
We probe a natural question if an instability arises  in a synchronized network of oscillators due to an evolving repulsive interaction, how the AD may originate and if it transits to the OD states? A direct transition from oscillatory to either AD or OD was reported
~\cite{PC Matthews:PRL_1990,Hou:2003,Koseska:PRE_2007,Zou:2009,Resmi:network,JJ Suarez:EPL_2009} in networks of oscillators, however, a transition from  AD to OD was not explored in networks of oscillators so far, until recently, this issue has been discussed briefly ~\cite{Banerjee} in a network of LS systems. Exploring the possibility of this transition is especially important in relation to testing robustness of the AD or HSS, in real systems, such as a cell signaling network \cite{kondor} under perturbation. A growth of inhibitory links may destabilize the HSS to transit to IHSS. It is a deeper question of understanding the dynamics of a  network how it is affected by repulsive interactions that may originate with time as a disease in a biological network~\cite{kondor}, a fault in a synchronized power grid~\cite{Motter}, an awareness campaign for arresting a spreading epidemics~\cite{Wu}. It is not an easy task to develop an understanding of this behavior in such networks of complex topologies.% and so far not addressed appropriately. 
\par In this paper,  instead we consider a simpler network  of globally coupled identical oscillators, limit cycle and  chaotic. We first  establish a synchrony of the network and then perturb it by adding a repulsive mean-field link $(N_p)$ between any two oscillators. We continue to add on the repulsive links to the other nodes and search for when the AD originates and transits to possible OD states. We mention that a repulsive mean-field coupling either in a global linear form \cite{franci} or a local nonlinear form \cite{kondor} was used earlier in realistic models. In our network model, we consider the linear repulsive mean-field links that spread into the local nodes only.
We have tested earlier \cite{CR Hens:PRE_2013} that AD emerges in a globally coupled network for additional repulsive links but fewer than the total number of nodes. However, the transition from AD to OD was not investigated there. We are now able to find the AD to OD transition which we  demonstrate here using both analytical and numerical techniques.
\section{Transition in global network of limit cycle oscillators}
%A small size network (N=20) using fewer repulsive links ($ N_p>\frac{N}{2}$) is first considered.
 \begin{figure}[h]
\includegraphics[height=!,width=7.5cm]{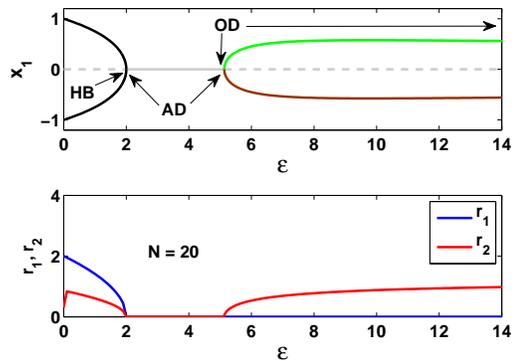} 
\caption{(Color online) AD and OD in globally coupled network of LS oscillators ($N=20$, $N_p=11$).  A bifurcation diagram plot in upper panel
(extrema of $x_{1,2}$ with coupling strength $\epsilon$) using MATCONT software ~\cite{Matcont:2003}, shows AD in gray line and OD in dark lines (brown and green lines). Lower panel plots $r_1$ (blue line) and $r_2$ (red line) with $\epsilon$ for $\omega = 3.0$. 
 } 
\label{fig:matcont_bif}
\end{figure}
\par  We start with the globally coupled network separated into two groups: a group of unperturbed nodes and a group of perturbed nodes by the repulsive links. %(say, GR:II).  
The first population with attractive diffusive coupling is,
%\begin{eqnarray}
\vspace{-0.15cm}
\begin{eqnarray}
\label{eq:sub_population_2}
\dot{\bf X_{\it l}} =f({\bf X_{\it l}}) + \frac{\epsilon}{N}\Gamma_1\sum_{j=1}^{N}({\bf X_j - X_{\it l}}).   
\end{eqnarray}
where $f:{{R}}^m\rightarrow {{R}}^m$, $\Gamma_1$ is a $m\times m$ matrix that includes the variables of the diffusive coupling and $l=p+1,...,N$.
The second population with perturbed nodes is described by,
\vspace{-0.15cm}
\begin{eqnarray}
\dot{\bf X_k} = f({\bf X_k}) + \frac{\epsilon}{N}\Gamma_1\sum_{j=1}^{N}({\bf X_j - X_k}) -\epsilon\Gamma_2({\bf X_k + X^*}.)
\vspace{-0.15cm}
\label{eq:sub_population_1}
\end{eqnarray}
where $\Gamma_2$ is a $m\times m$ matrix that includes the variables of the repulsive coupling; $k=1,2,...,p$ is the node number and ${\bf X^*}$  represents any node of the first population. The $\epsilon\Gamma_2 ({\bf X_k + X^*})$ represents the additional repulsive link for a positive $\epsilon$.
As example, we consider the LS system %$\dot{z_i}=[1+iw-{{z_i}}^2]z_i$, where $z_i=x_i+iy_i$, 
representing the dynamics of each node of the network. Two groups of LS oscillators are written as, 
\ \vspace{-0.05cm}
\begin{eqnarray}
\label{eq:reduced_population_2}
\dot{x_{\it l}}&=&[1 - (x_{\it l}^2 + y_{\it l}^2)]x_{\it l} - \omega y_{\it l}+\frac{\epsilon}{N}\sum_{j=1}^{N}(x_j - x_{\it l}),\\
\dot{y_{\it l}}&=&[1 - (x_{\it l}^2 + y_{\it l}^2)]y_{\it l} + \omega x_{\it l},\nonumber
\end{eqnarray}\vspace{-0.25cm}
and 
\begin{eqnarray}
\label{eq:reduced_population_1}
\dot{x_k}&=&[1 - (x_k^2 + y_k^2)]x_k - \omega y_k+
 \frac{\epsilon}{N}\sum_{j=1}^{N}(x_j - x_k), \\
\dot{y_k}&=&[1 - (x_k^2 + y_k^2)]y_k + \omega x_k -\epsilon(y_k + y_N).\nonumber
\vspace{-.5cm}
\end{eqnarray} 
The attractive global coupling is applied through the $x$-variable  while the repulsive link is applied via the y-variable. The network is divided into $p-$ and $q-$ sub-populations such that $p$ number of oscillators are connected by 
 repulsive links in addition to the global attractive coupling and $q = N - p$ number of oscillators are not perturbed by any repulsive link.  
Now consider a smaller network of identical LS oscillators of $N = 20$ (natural frequency $\omega = 3.0$) and add $N_p=11$ repulsive links and, assume a symmetric coupling, $\epsilon_1 = \epsilon_2 = \epsilon$ for further reduction of complexity. 
We make a bifurcation analysis  using the MATCONT software~\cite{Matcont:2003} as well as the numerical simulations which are shown in Figs. \ref{fig:matcont_bif}(a) and (b) respectively. 
Figure \ref{fig:matcont_bif}(a) shows the onset of AD via HB at $\epsilon = 2$ when a stable limit cycle (LC) in solid black line becomes a stable HSS at equilibrium origin (gray line). The AD transits to OD at a larger coupling ($\epsilon=5.2$) via the PF bifurcation as usually found \cite{A Koseska:Phys Rep_2013} in  two diffusively coupled LS oscillators with a mismatch, which gives rise to two stable branches of the OD (online brown and green line) and a coexisting unstable equilibrium origin (gray dashed line). 
For  further characterization of the AD and OD regimes, we define two order parameters, $r_1$ and $r_2$, \
\begin{equation}
r_1=\frac{1}{N}\sum_{i=1}^{N}|\max(x_i(t))-\min(x_i(t))|. %\nonumber,
\label{eq:r_1}\nonumber
\end{equation}
and 
\begin{equation}
r_2 = \frac{1}{N}\sum_{i=1}^{N}<\sqrt{((x_i(t) - x_1(t))^2 + (y_i(t) - y_1(t)^2)^2}>. %\nonumber,
\label{eq:r_2}\nonumber
\end{equation}
where $<\cdot>$ denotes the time average over all the nodes of the network. We make a look-up table for characterization of the LC, AD and OD states explicitly by the $r_1$ and $r_2$ measures plotted in Fig. \ref{fig:matcont_bif}(b),
\begin{center}
    \begin{tabular}{| l | l | l | l |}
    \hline
     & LC & HSS & IHSS \\ \hline
   $ r_1$ & $\neq0$ & 0 & 0  \\ \hline
    $r_2$ & $\neq0 $& 0 &$\neq0$ \\
       \hline
    \end{tabular}
\end{center}
Figure \ref{fig:matcont_bif}(b) confirms the AD and OD regimes with varying coupling strengths $\epsilon$. $r_1 \rightarrow 0$ for a critical coupling $\epsilon>=2.0$ that marks the onset of AD and it remains zero for larger $\epsilon$ values. The $r_2$ measures the average distance of all the nodes in the network from the first node ($i=1$), and signifies an AD (HSS) for $2<\epsilon<5.2$. For $\epsilon>5.2$, the OD (IHSS) sets in the network when only $r_2 \neq 0$. 
\begin{figure} [h]
\includegraphics[height=!,width=9cm]{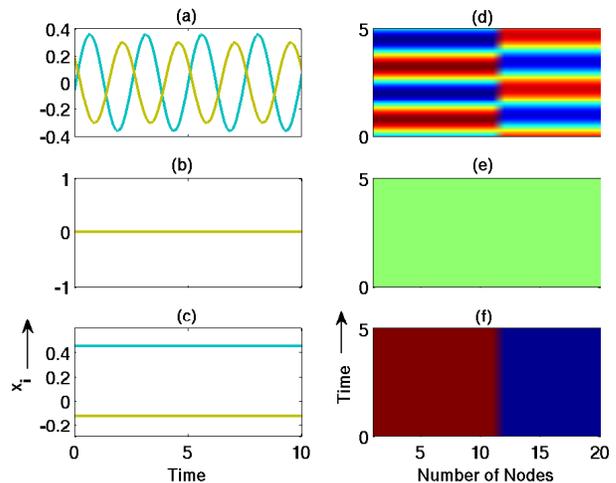}
\caption{(Color online) Network of LS oscillators (N=20, $N_p=11$). Superimposed time series of $x_i$ of all (i=20) nodes at left column, (a) two clustered states in out-of-phase oscillatory mode for $\epsilon=1.8$, (b) one HSS line for $\epsilon=4.0$ and (c) two IHSS lines for $\epsilon=6.0$. Time-series $x_i$ for all the nodes in spatio-temporal  plots in (d), (e) and (f) at right column that {\bf corroborates} their immediate left plots.}
\label{fig:TS_STP_20nodes}
\end{figure}
Figures \ref{fig:TS_STP_20nodes}(a), \ref{fig:TS_STP_20nodes}(c) and \ref{fig:TS_STP_20nodes}(e) 
show the superimposed time-series of  $x_i$ of all the nodes ($N = 20$) of the network for three different coupling strengths where distinctly different dynamical regimes are identified. 
Figure \ref{fig:TS_STP_20nodes}(a) for $\epsilon = 1.8$ shows that all the nodes are oscillatory but synchronized to form two clusters of dynamical units in an out-of-phase mode. For a larger $\epsilon=4.0$, all the nodes converge to a single cluster state basically indicating a stable HSS in Fig. \ref{fig:TS_STP_20nodes}(c), while two stable HSS states arrive at $\epsilon=6.0$ as shown in Fig. \ref{fig:TS_STP_20nodes}(e).
Figures \ref{fig:TS_STP_20nodes}(b), \ref{fig:TS_STP_20nodes}(d) and \ref{fig:TS_STP_20nodes}(f) plot the respective spatio-temporal evolution of all the $N=20$ nodes, which reconfirm the signatures of the panels at immediate left:  
two oscillatory clusters in Fig. \ref{fig:TS_STP_20nodes}(b), single HSS in Fig. \ref{fig:TS_STP_20nodes}(d) and two IHSSs in Fig. \ref{fig:TS_STP_20nodes}(f).
\par 
\begin{figure}[h]
\includegraphics[height=6cm,width=7cm]{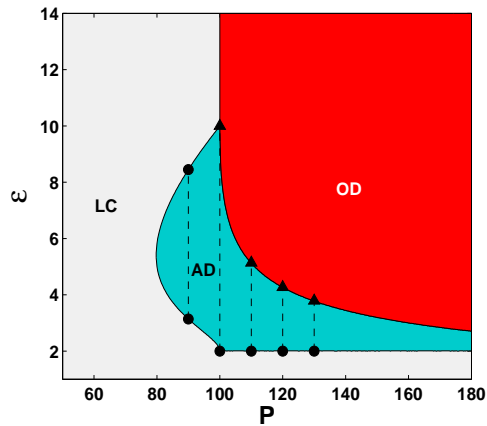}
\caption{Phase diagram in $\epsilon-p$ plane of the reduced model of the network of LS units ($N=200$, $\omega = 3.0$). For different values of $p$, $k = 1, 2,\dots, p$ and $l = p+1, p+2, \dots, N$. Limit cycle (LC), AD and OD regions are delineated. The gray, light gray (cyan) and black (red) regions denote the LC, AD and OD states respectively. Black circles and triangles represent AD  and  OD respectively as obtained from a direct numerical simulation of the full model with $N=200$. Numerical results perfectly matches with the boundaries of LC to AD  and the AD to OD transitions. } 
\label{fig:LC_AD_OD}
\end{figure}
{\bf Noteworthy} that we start with a globally synchronized coupled network where the synchronization manifold is given by ${\bf x_1 = x_2 = ... = x_N}$ for a certain threshold of $\epsilon$. A fraction (say, $p$ oscillators) of the total population is then perturbed by the additional repulsive mean-field links which splits the synchronized network into two populations (two oscillatory clusters) in an out-of-phase mode. More explicitly, we can express this splitting by two out-of-phase manifolds,
${\bf x_1 = x_2 = \cdot\cdot\cdot = x_p}$;
and ${\bf x_{p+1} = x_{p+2} = \cdot\cdot\cdot = x_N}$
where $q = N - p$. We make an assumption here that a repulsive mean field link does not connect two oscillators from the same population. 
Hence, one can easily reduce the eqs.~(\ref{eq:reduced_population_2}) and ~(\ref{eq:reduced_population_1}) to represent two populations of oscillators by their dynamical variables $(X_{1,2}, Y_{1,2})$, 
\begin{eqnarray}
\label{eq:reduced_popu_final}
\dot{X_2}&=&[1 - (X_2^2 + Y_2^2)]X_2 - \omega Y_2+  \frac{p\epsilon}{N}(X_1-X_2), \\
\dot{Y_2}&=&[1 - (X_2^2 + Y_2^2)]Y_2 + \omega X_2, \nonumber\\
\dot{X_1}&=&[1 - (X_1^2 + Y_1^2)]X_1 - \omega Y_1+  \frac{q\epsilon}{N}(X_2-X_1),  \\
\dot{Y_1}&=&[1 - (X_1^2 + Y_1^2)]Y_1 + \omega X_1 -\epsilon(Y_1+Y_2). \nonumber
\end{eqnarray}
%Here variables of two oscillators are described by $(X_1,Y_1)$ and $X_2,Y_2$. 
The reduced system has a trivial equilibrium origin. The stability of the origin is determined by the {\it Jacobian} $J$ of the reduced model,
\[J= \left(\begin{array}{cccc}
\frac{p\epsilon}{N} & 0 & 1-\frac{p\epsilon}{N} & -\omega \\
0 & 0 & \omega & 1 \\
1- \frac{q\epsilon}{N} & -\omega & \frac{q\epsilon}{N} & 0 \\
\omega & 1-\epsilon & 0 & -\epsilon  \\
\end{array} \right)\] 
We make an eigenvalue analysis of the $J$ using the MATLAB tool and plot a phase diagram in the $\epsilon-p$ plane in Fig. \ref{fig:LC_AD_OD}. Regimes of LC, AD and OD are clearly delineated by white, gray (cyan online) and dark gray (red online) regions with clear boundary lines. In the LC region, the origin is unstable and  two pairs of complex conjugate eigenvalues exist with positive real part. A negative sign change of the largest real part of the complex conjugate eigenvalue across the LC to AD boundary line, indicates a transition to the AD regime via HB: a stabilization of the  equilibrium origin. On the other hand, a pure real positive eigenvalue appears across the AD to OD boundary where the stable origin becomes unstable. We note that we have already identified the origin of two new stable equilibrium points at this boundary. Each of the vertical dashed lines shows a transition from LC to OD via an intermediate AD for increasing $\epsilon$ values and a fixed number of repulsive links. If the repulsive links are smaller than $p=100$ for our example network of $N=200$, no transition to OD is possible and it is clearly understood when we look at the  meeting point of the LC, AD and OD boundaries for $\epsilon=10, p=100$. We have checked that this is valid for larger $N$ values. At least $N_p=p=N/2$ repulsive links are necessary to induce a transition from LC to OD state via AD or a direct transition to OD state. This information guides us to start with $N_p=N/2$ repulsive links to draw the bifurcation diagram in Fig. 1. However, the AD can appear for lower coupling strength and lower number of repulsive links, which we have reported earlier \cite{CR Hens:PRE_2013}. 
We verify the analytical results by simulating the full model with N=200 units of coupled (both attractive and repulsive links) LS oscillators and denote the onset of AD by black circles and OD by black triangles which exactly fall on the respective borderlines. They are determined by using the two order parameters ($r_1, r_2$).

\section{Network of chaotic oscillators}
\par
\begin{figure}[h]
\includegraphics[height=6cm,width=8cm]{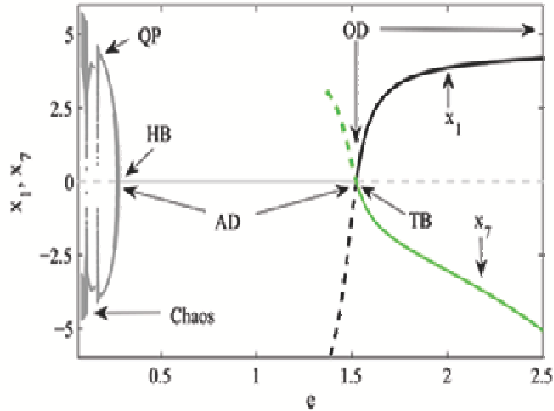}
\caption{Globally coupled network of R\"ossler oscillators ($N=10, N_p=6$) with $k = 1, 2,\dots, 6$ and $l = 7, 8, 9, 10$. Extrema of $x_1$ and $x_7$ are plotted with $\epsilon$. Solid and dashed light gray lines respectively represent the stable and unstable AD states. Stable AD state (solid light gray line) appears via inverse HB at $\epsilon = 0.27$. AD  transits to OD via TB at $\epsilon = 1.52$. Parameter values are: $a = 0.36$, $b = 0.4$, $c = 4.5$.} 
\label{fig:OS_AD_OD}
\end{figure}

 We extend our observation on AD-OD transition to a network of chaotic R\"ossler oscillators. We separate all the globally coupled R\"ossler oscillators once again into two sub-populations, one perturbed by repulsive links and another unperturbed,
\begin{eqnarray}
\label{eq:ross_popu_1}
\dot{x_{\it l}}&=& - y_{\it l}- z_{\it l} + \frac{\epsilon}{N}\sum_{j=1}^{N}(x_j - x_{\it l}),\\
\dot{y_{\it l}}&=&x_{\it l} + ay_{\it l},\nonumber\\
\dot{z_{\it l}}&=&bx_{\it l} + z_{\it l}(x_{\it l} -c).\nonumber
\end{eqnarray}\vspace{-0.25cm}
and 
\begin{eqnarray}
\label{eq:ross_popu_2}
\dot{x_k}&=&- y_{\it k}- z_{\it k} + \frac{\epsilon}{N}\sum_{j=1}^{N}(x_j - x_{\it k}),\\
\dot{y_k}&=&x_{\it k} + ay_{\it k}-\epsilon(y_k + y_N),\nonumber\\
\dot{z_{\it k}}&=&bx_{\it k} + z_{\it k}(x_{\it k} -c).\nonumber
\vspace{-0.25cm}
\end{eqnarray}
where $k=1,2....p$ and $l=p+1,p+2....N$. 
To investigate the origin of AD and its transition to OD, we first take a network of $N = 10$ and $N_{\it l} = 6$.  
We perform the bifurcation analysis of the coupled network  using the MATCONT. The bifurcation diagram in Fig.~\ref{fig:OS_AD_OD} shows that for very low $\epsilon$, network displays a chaotic dynamics. As $\epsilon$ value is increased, quasiperiodic solutions appear followed by the AD/HSS state via inverse HB at $\epsilon = 0.27$. AD/HSS state observed in the range  $0.27 \leq \epsilon \leq 1.52$ and it transits to OD (two stable steady states $x_1$ in black line and $x_7$ in green line) via transcritical bifurcation (TB) at $\epsilon=1.52$.
The mechanism process of transition from AD to OD is thus found different from PF bifurcation, however, similar to the process reported earlier \cite{CR Hens:PRE_2014} for two chaotic oscillators. 
Now we verify the results, once again, using numerical simulation of a network of R\"ossler oscillators of size $N = 10$ and $N_p = 6$ that two synchronized clusters in out-of-phase mode, indeed, appear for this system too before the transition to the AD state. 
Once confirmed about this 2-cluster effect, we can use a reduced system of the large network of R\"ossler oscillators where we introduce the variable $X_{1,2}$, $Y_{1,2}$ and $Z_{1,2}$,
\begin{eqnarray}
\label{eq:reduced_popu_final}
\dot{X_2}&=&-Y_2 - Z_2 +  \frac{p\epsilon}{N}(X_1-X_2), \\
\dot{Y_2}&=&X_2+aY_2, \nonumber\\
\dot{Z_2}&=&bX_2+Z_2(X_2-c), \nonumber\\
\dot{X_1}&=&-Y_1 - Z_1 +  \frac{q\epsilon}{N}(X_2-X_1),  \\
\dot{Y_1}&=& X_1+aY_1 - \epsilon(Y_1+Y_2), \nonumber\\
\dot{Z_1}&=& bX_1+Z_1(X_1-c).\nonumber
\end{eqnarray}\label{reduced_rossler}

\begin{figure}[h]
\includegraphics[height=6cm,width=7cm]{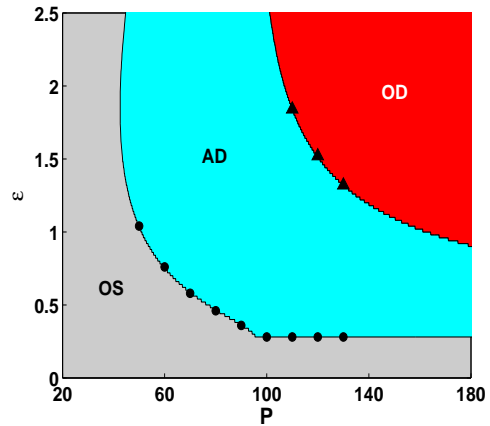}
\caption{(Color online) Phase diagram in $\epsilon-p$ plane of the reduced model of the network of R\"ossler units ($N=200$, $a=0.36$, $b=0.4$, $c=4.5$). For different values of $p$, $k = 1, 2,\dots, p$ and $l = p+1, p+2, \dots, N$. OS, AD and OD regions are delineated by gray, light gray (cyan) and black (red) lines respectively. Solid black circles and triangles represent AD  and  OD respectively from numerical simulation of the full system. For a fixed number of repulsive links more than N/2, an AD-OD transition is possible for increasing coupling strength $\epsilon$. On the other hand, for a fixed coupling strength $\epsilon$, such a transition can be incurred by increasing the number of repulsive links.} 
\label{fig:rossler_AD_OD}
\end{figure}

The attractive coupling is applied via the $x-$variable and the repulsive interaction is introduced in the $y-$variable. 
A phase diagram (Fig.~\ref{fig:rossler_AD_OD}) is constructed using eigenvalue analysis of the reduced system of the $N = 200$ network. The reduced system has a trivial equilibrium origin whose stability is analyzed to draw this phase diagram. In Fig.~\ref{fig:rossler_AD_OD}, three distinct regimes namely oscillatory (OS), AD and OD with clear boundaries are identified using eigenvalue analysis similar to what is observed for the LS network. Note that the OS regime consists of three different oscillatory regimes, chaotic, quasiperiodic and periodic, which we do not elaborate here since it is beyond the target of this work. We now perform a numerical simulation of the full network of $N = 200$ for different number of negative links and $\epsilon$ values. The AD and OD points obtained from the numerical simulations are plotted with black dots and triangles respectively in the Fig.~\ref{fig:rossler_AD_OD} and they perfectly fall on the respective AD and OD boundaries. 
\section{conclusions}
\par In summary, we observed an emergence of AD and its transition to OD in networks of synchronized oscillators when perturbed by repulsive interaction. A simple network of globally coupled oscillators was considered. Both analytical and numerical approaches were adopted to identify the AD and OD regimes and their transitions. A smaller size network was first taken to recognize the AD and the OD regimes using the MATCONT software and then v
erified the result using a direct numerical simulation of larger networks. We identified that a synchronous regime of two clusters in out-of-phase mode first appears for increasing coupling strength followed by a transition to the AD state for a larger coupling and then to the OD state. Based on this clustering information, we proposed a reduced model of the large network that allowed us to make an eigenvalue analysis to delineate the OS, AD and OD regimes in a phase plane of the number of repulsive links and the coupling strength. Secondly, we verified the AD and OD boundaries by numerically simulating the full network for selected number repulsive links and coupling strength that perfectly matched with the analytically drawn boundary lines. We checked the result for networks of both limit cycle LS systems and chaotic R\"ossler systems. However, the mechanism of transition from AD to OD for a network of limit cycle system is PF bifurcation as usual \cite{CR Hens:PRE_2013} while it is TB for a network of chaotic oscillators as reported for a simple model of two chaotic oscillators \cite{CR Hens:PRE_2014}. 
The fundamental nature of the transitions did not change with an increase of the size of the network. We make a future plan to study on more realistic complex dynamical networks such  as the power-grid using simple model \cite{Filatrella}

 to investigate how they could be affected by additional repulsive interaction  as a local fault evolves in time and spread into the network to induce a death like situation analogous to the AD regime. Further we investigate the effect of repulsive or inhibitory links in a cell signaling network to test how robust a HSS and if it breaks into IHSSs. 

\begin{acknowledgments}
S.K.D. and C.R.H. acknowledge support by the CSIR Emeritus scientist scheme.
\end{acknowledgments}

%\begin{thebibliography}{9}\label{sec:TeXbooks}%
%\bibitem{Note1}
%For help regarding the installation of this software and its use, please send email to \href{mailto:tex@aip.org}{tex@aip.org}.
%%
%\bibitem{Note2}
%Available with the \revtex\ distribution, see \url{http://authors.aps.org/revtex4/}.
%%
%\bibitem[Lamport(1996)]{LaTeXman} 
%L. Lamport, 
%\emph{\LaTeX\, a Document Preparation System} 
%(Addison-Wesley, Reading, MA, 1996).
%%
%\bibitem[Goossens(1994)]{Compan} 
%M. Goosens, F. Mittelbach, and A. Samarin, 
%\emph{The \LaTeX\ Companion} 
%(Addison-Wesley, Reading, MA, 1994).
%%
%\bibitem[Knuth(1986)]{TeXbook} 
%D. E. Knuth, 
%\emph{The \TeX book} 
%(Addison-Wesley, Reading, MA, 1986). 
%%
%\bibitem[Kopka(1995)]{Guide} 
%H. Kopka and P. Daly, 
%\emph{A Guide to \LaTeXe} 
%(Addison-Wesley, Reading, MA, 1995).
%%
%\bibitem[Goossens(1997)]{CompanG} 
%M. Goossens, S. Rahtz, and F. Mittelbach, 
%\emph{The \LaTeX\ Graphics Companion} 
%(Addison-Wesley, Reading, MA, 1997).
%%
%\bibitem[Rahtz(1999)]{CompanW} 
%S. Rahtz, M. Goossens \emph{et al.},
%\emph{The \LaTeX\ Web Companion} 
%(Addison-Wesley, Reading, MA, 1999).
%%
%\end{thebibliography}
 
\end{document}